\documentclass[a4paper]{article}
\usepackage{amssymb}
\usepackage{amsmath}
\usepackage{latexsym}
\usepackage{mathrsfs}
\usepackage{cancel}

\providecommand{\pacs}[1]{PACS numbers : #1}
\providecommand{\keywords}[1]{Keywords : #1}

\begin{document}

\title{Maxwell's equations in the context of the Fock transformation and the magnetic monopole}
\author { N.~Takka \footnote{E-mail:
{\tt takkanaimi@yahoo.fr }}
\ A.~Bouda \footnote{E-mail:
{\tt bouda\_a@yahoo.fr}}
\ and T.~Foughali \footnote{E-mail:
{\tt fougto\_74@yahoo.fr }}\\
Laboratoire de Physique Th\'eorique, Facult\'e des Sciences Exactes,\\
Universit\'e de Bejaia, 06000 Bejaia, Algeria\\}

\maketitle

\begin{abstract}
In the $R$-Minkowski space-time, which we recently defined from an appropriate deformed Poisson brackets that reproduce the Fock coordinate transformation, we derive an extended form for Maxwell's equations by using a generalized version of Feynman's approach. Also, we establish in this context the Lorentz force. As in deformed special relativity, modifying the angular momentum in such a way as to restore the $R$-Lorentz algebra generates the magnetic Dirac monopole.
\end{abstract}



\newpage


\section{Introduction}
Fock's nonlinear relativity is characterized by the following coordinate transformation \cite{Fock}
\begin{equation}
t^{\prime }=\frac{\gamma(t-ux/c^2)}{\alpha_R},\quad x^{\prime }=\frac{
\gamma(x-ut)}{\alpha_R},\quad y^{\prime }=
                                         \frac{y}{\alpha_R},\quad z^{\prime}
                                                                           =\frac{z}{\alpha_R},
\end{equation}
where
$
\gamma =(1-u^{2}/c^{2})^{-1/2}
$,
$
{\alpha _{R}}=1+\left[(\gamma -1)ct-\gamma xu/c\right]/R
$
and $R$ being the universe radius.
This transformation was recently reproduced from an appropriate deformed Poisson brackets
\cite{BF} by using an analogous procedure as in \cite{Ghosh-Pal}, where the coordinate
transformation of Deformed Special Relativity \cite{Ameliano1, Ameliano2, Mag-Smol1, Mag-Smol2} is
recovered. It is shown then that in addition to the constant $c$ which represents the speed
of light at the limit $R\rightarrow\infty$, transformation (1) keeps also invariant the universe
radius. It defines the so-called $R$-Minkowski spacetime. Contrary to earlier versions of Fock's
nonlinear relativity, this approach involving deformed Poisson brackets generated a transformation
of the momentum with which the contraction $p_{\mu}x^{\mu}$ is an invariant, allowing then a coherent
description of the plane waves in the context of the Fock transformation. After first quantization,
these Poisson brackets were replaced by the following commutations relations
\begin{align}
\left[x^\mu,x^\nu\right] & = 0, \\
\left[x^\mu,p^\nu\right] & = - i \hbar \eta^{\mu\nu} + \frac{i \hbar}{R}\eta^{0\nu}x^\mu, \\
\left[p^\mu,p^\nu\right] & = -\frac{i \hbar}{R}[p^\mu\eta^{0\nu}-p^\nu\eta^{\mu0}],
\end{align}
constituting the phase space algebra of the $R$-Minkowski spacetime \cite{FB1}.
Here $\eta^{\mu\nu} = (+1,-1,-1,-1)$ and $\mu ,\nu =0,1,2,3$. The above $R$-algebra was completed by
commutation relations involving angular momentum components,
\begin{equation}
\textbf{J}^{\mu \nu }=x^{\mu }p^{\nu }-x^{\nu }p^{\mu },
\end{equation}
which satisfy the following relations
\begin{align}
\left[x^{\mu}, \textbf{J}^{\nu\lambda}\right] & = -i\hbar \left(\eta^{\mu\lambda}x^{\nu}-\eta^{\mu\nu}x^{\lambda}\right) +
                                       \frac{i\hbar}{R}x^{\mu} \left(\eta^{0\lambda}x^{\nu}-\eta^{0\nu}x^{\lambda}\right), \\
\left[p^{\mu}, \textbf{J}^{\nu\lambda}\right] & =  i\hbar \left(\eta^{\nu\mu}p^{\lambda}-\eta^{\lambda\mu}p^{\nu}\right)+
                                       \frac{i\hbar}{R}p^{\mu} \left(\eta^{\nu 0}x^{\lambda}-\eta^{\lambda 0}x^{\nu}\right),\\
\left[\textbf{J}^{\mu\nu}, \textbf{J}^{\lambda\sigma}\right] & = i\hbar \left(\eta^{\nu\lambda}\textbf{J}^{\mu\sigma}-
                                                   \eta^{\nu\sigma}\textbf{J}^{\mu\lambda}
                                              -\eta^{\mu\lambda}\textbf{J}^{\nu\sigma}+\eta^{\mu\sigma}\textbf{J}^{\nu\lambda}\right).
\end{align}
The corresponding first Casimir is constructed and allowed the derivation of the Klein-Gordon
equation which turned out to be the Klein-Gordon equation in de Sitter Space given by its conformal
metric \cite{FB1}. This result established a correspondence between the $R$-Minkowski spacetime and de
Sitter's spacetime. Also this correspondence was confirmed since Dirac's equation was derived in
$R$-Minkowski spacetime \cite{FB2} and it turned out to be exactly the Dirac equation in the
conformally de Sitter spacetime.

Elsewhere, Dyson \cite {Dyson} claimed that Feynman derived Maxwell's equations and the Lorentz
force from Newton's law of motion and the commutator between position and velocity.
Tanimura \cite{Tanimura} formulated the relativistic version of Feynman's derivation by
starting from
\begin{align}
\left[x^\mu,x^\nu \right]&=0, \\
\left[x^\mu,\dot{x}^\nu \right]&=-\frac{i\hbar}{m}\eta^{\mu\nu}, \\
m\ddot{x}^{\mu} &= F^{\mu}(x,\dot{x})
\end{align}
and by defining the electromagnetic tensor as
\begin{equation}
F^{\mu\nu} \equiv -\frac{m^{2}}{i\hbar q} \left[\dot{x}^{\mu},\dot{x}^{\nu} \right],
\end{equation}
where $x^{\mu}=x^{\mu}(\tau)$, $\tau$ is a parameter with a dimension of time, the dot refers to the
derivative with respect to $\tau$ and $q$ is the charge of a particle of mass $m$. He derived the relations
\begin{align}
0 & = \partial^{\lambda}F^{\mu\nu}+ \partial^{\nu}F^{\lambda\mu}+\partial^{\mu}F^{\nu\lambda}, \\
F^{\mu} & = qF^{\mu\nu} \dot{x}_{\nu},
\end{align}
which represent respectively the first group of Maxwell's equations and the Lorentz force.
Later, B\'{e}rard et al \cite{BGM1,BGM2,BM} developed this approach and showed how the
magnetic monopole appears as a consequence of the restoration of the Lorentz algebra symmetry. 
Also, Harikumar et al \cite{H1,H2} established the Deformed Special Relativistic version of 
this formulation, up to first order in the deformation parameter, by using the commutation 
relations of the $\kappa$-Minkowski spacetime \cite{Ghosh-Pal}. In this extension, it is shown 
that the $\kappa$-deformed Maxwell equations preserve the electric-magnetic
duality and predict the manifestation of the magnetic monopole.

In this paper, we will generalize the use of the Feynman method to Fock's nonlinear relativity.
In section 2, we will establish, up to first order in the deformation, the deformed Maxwell
equations and Lorentz force in the $R$-Minkowski spacetime. In section 3, we restore the
$R$-Lorentz symmetry by modifying the momentum generators and therefore we prove the presence of
the Dirac magnetic monopole. Section 4 is devoted to conclusion.

\section{Maxwell's equation in $R$-Minkowski spacetime}

In order to establish Maxwell's equations in the context of Fock's nonlinear relativity,
let us first search for the analogous of relations (10) and (12) in the framework of
$R$-Minkowski spacetime. For this purpose, let us recall that it is shown in \cite{BF} that
the following canonical variables
\begin{align}
X^{\mu} & = \left(1-\frac{x^0}{R}\right)^{-1}x^{\mu}  , \\
P^{\mu} & =  \left(1-\frac{x^0}{R}\right)p^{\mu}
\end{align}
obey to the laws of special relativity. Thus if we substitute the Poisson brackets by commutators, we have
\begin{align}
\left[ X^{\mu}, X^{\nu}\right] & =   0, \\
\left[ X^{\mu}, P^{\nu}\right] & =   -i \hbar \eta^{\mu\nu},  \\
\left[ P^{\mu}, P^{\nu}\right] & =   0.
\end{align}
At first order in $1/R$, relation (15) gives
\begin{equation}
X^{\mu} = \left(1+\frac{x^0}{R}\right)x^{\mu}
\end{equation}
and then
\begin{equation}
\dot{X}^{\mu} = \dot{x}^{\mu} + \frac{1}{2R} \left(\dot{x}^{0}x^{\mu} +
                                  x^{\mu}\dot{x}^{0} + x^{0}\dot{x}^{\mu} + \dot{x}^{\mu} x^{0} \right),
\end{equation}
where the symmetrization operation was performed.

Let us begin by the case where the electromagnetic field is absent. Using the fact that
\begin{equation}
P^{\mu}=m\dot{X}^{\mu},
\end{equation}
relations (18) and (19) allow to write
\begin{align}
\left[ X^{\mu}, \dot{X}^{\nu} \right] & =  - \frac{i \hbar}{m} \eta^{\mu\nu} , \\
\left[ \dot{X}^{\mu}, \dot{X}^{\nu} \right] & =   0.
\end{align}
Using expressions (20) and (21) in (23), we obtain
\begin {eqnarray}
\left[ x^{\mu}, \dot{x}^{\nu} \right] +
\frac{1}{R} \left\{
                   \frac{3}{2} x^{0} \left[ x^{\mu}, \dot{x}^{\nu} \right] +
                   \frac{1}{2} \left[ x^{\mu}, \dot{x}^{0} \right] x^{\nu}
                                                                            \right.
                  \hskip30mm&& \nonumber \\
\left.
      + \frac{1}{2} x^{\nu} \left[ x^{\mu}, \dot{x}^{0} \right]
      + \frac{1}{2} \left[ x^{\mu}, \dot{x}^{\nu} \right] x^{0}
      + \left[ x^{0}, \dot{x}^{\nu} \right] x^{\mu}
                                                               \right\}
                 = - \frac{i \hbar}{m} \eta^{\mu\nu}.
\end {eqnarray}
The zeroth order solution of this last equation is obtained by taking the limit $R\rightarrow \infty$
\begin{equation}
\left[ x^{\mu}, \dot{x}^{\nu} \right]_{(0)}  =  - \frac{i \hbar}{m} \eta^{\mu\nu}.
\end{equation}
At first order, all the commutators between braces in (25) can be replaced by their
value of zeroth order since they are preceded by the factor $1/R$. Then, we deduce that
\begin{equation}
\left[ x^{\mu}, \dot{x}^{\nu} \right]  =  - \frac{i \hbar}{m} \eta^{\mu\nu} +
   \frac{i \hbar}{mR} \left( 2x^{0} \eta^{\mu\nu} + x^{\nu}\eta^{\mu0} + x^{\mu}\eta^{0\nu} \right).
\end{equation}
In the same way, the use of (21) in (24) leads to
\begin{equation}
\left[ \dot{x}^{\mu}, \dot{x}^{\nu} \right]  =
               \frac{i \hbar}{Rm} \left( \dot{x}^{\mu}\eta^{0\nu} -  \dot{x}^{\nu}\eta^{\mu 0} \right).
\end{equation}
Contrary to the special relativistic case, we see that this commutator has a non-zero value.
By using (16), we can check that this result is compatible with relation (4).

Let us now move on to the case where the electromagnetic field is present.
As in special relativistic case, the commutator $\left[ X^{\mu}, \dot{X}^{\nu} \right]$ keeps the
same expression in the absence or in the presence of the electromagnetic field. In fact, taking into
account relation (17), even if we use in (18) instead of (22) the following expression
\begin{equation}
P^{\mu}=m\dot{X}^{\mu} + \frac{q}{c} A^{\mu}(X),
\end{equation}
we reproduce (23), $A^{\mu}$ being the four-potential. For the same reason, if we follow
the same procedure as above to determine the commutator $\left[ x^{\mu}, \dot{x}^{\nu} \right]$
in presence of the electromagnetic field, we reproduce expression (27).

Concerning the commutator $\left[ \dot{x}^{\mu}, \dot{x}^{\nu} \right]$, definition
(12) allows us to observe the following properties in the special relativistic case :
\begin{enumerate}

\item The electromagnetic tensor is antisymmetric : $F^{\mu\nu}=-F^{\nu\mu}$.

\item In the absence of the electromagnetic field, the commutator
$\left[ \dot{x}^{\mu}, \dot{x}^{\nu} \right]$ takes a vanishing value,
what is compatible with the fact that $\left[ p^{\mu}, p^{\nu} \right]=0$.

\item The electromagnetic tensor does not depend on $\dot{x}^{\lambda}$ since
$\left[F^{\mu\nu} , x^{\lambda} \right]=0 $ .

\end{enumerate}
\noindent
In the context of Fock's nonlinear relativity, definition (12) does not work since the second and the third
of the above properties are not satisfied. However, if we define
\begin{equation}
F^{\mu\nu} \equiv -\frac{m^{2}}{i\hbar q} \left[\dot{x}^{\mu},\dot{x}^{\nu} \right]
       + \frac{m}{qR} \left( \dot{x}^{\mu}\eta^{0\nu} -  \dot{x}^{\nu}\eta^{\mu 0} \right) \nonumber,
\end{equation}
meaning that
\begin{equation}
\left[\dot{x}^{\mu},\dot{x}^{\nu} \right] = \frac{i\hbar}{Rm} \left(\eta^{0\nu}\dot{x}^{\mu} - \eta^{\mu 0}\dot{x}^{\nu} \right)
                                            -\frac{i\hbar q}{m^{2}}F^{\mu\nu},
\end{equation}
we observe that $F^{\mu\nu}$ keeps its antisymmetric property and that the commutator
$\left[\dot{x}^{\mu},\dot{x}^{\nu} \right]$ reduces in the absence of the electromagnetic field
to expression (28) established above. We can also check that $F^{\mu\nu}$ does not depend on
$\dot{x}^{\lambda}$. In fact, substituting expression (30) in the following Jacobi identity
\begin{equation}
\left[\left[\dot{x}^{\mu},\dot{x}^{\nu} \right], x^{\lambda}\right] =
                     - \left[ \left[x^{\lambda}, \dot{x}^{\mu} \right], \dot{x}^{\nu} \right]
                     - \left[\left[\dot{x}^{\nu}, x^{\lambda} \right], \dot{x}^{\mu}\right] \nonumber,
\end{equation}
we deduce that
\begin {eqnarray}
\left[ F^{\mu\nu}, x^{\lambda} \right] = \frac{m^{2}}{i \hbar q}
                                                \left\{ \left[ \left[x^{\lambda}, \dot{x}^{\mu} \right], \dot{x}^{\nu} \right]
                                                        + \left[\left[\dot{x}^{\nu}, x^{\lambda} \right], \dot{x}^{\mu} \right]
                                                 \right\}
                                                 \hskip20mm&& \nonumber \\
                                            + \frac{m}{qR} \left\{ \eta^{0\nu} \left[\dot{x}^{\mu}, x^{\lambda} \right]
                                            -  \eta^{\mu 0} \left[\dot{x}^{\nu}, x^{\lambda} \right] \right\} .
\end {eqnarray}
With the use of (27), it is easy to show that $\left[F^{\mu\nu} , x^{\lambda} \right]=0 $, meaning that
$F^{\mu\nu} $ does not depend on $\dot{x}^{\lambda}$.

In order to establish the first group of Maxwell equations, as in \cite{Tanimura} our starting point is
the following Jacobi identity
\begin{equation}
\left[\dot{x}^{\mu},\left[\dot{x}^{\nu}, \dot{x}^{\lambda}\right] \right]
                + \left[ \dot{x}^{\lambda}, \left[\dot{x}^{\mu} , \dot{x}^{\nu} \right] \right]
                            + \left[\dot{x}^{\nu}, \left[\dot{x}^{\lambda} , \dot{x}^{\mu}\right] \right] = 0.
\end{equation}
Relations (27) and (2) indicate that $\left[x^{\lambda},\left[\dot{x}^{\nu}, x^{\alpha}\right] \right] =0 $. This
allows to write
\begin{equation}
\left[\dot{x}^{\mu}, F^{\nu\lambda}\right] =
      \frac{\partial F^{\nu\lambda}}{\partial x^{\alpha}} \left[\dot{x}^{\mu}, x^{\alpha}\right].
\end{equation}
Using relation (30) in (32) and taking into account (33), we deduce that
\begin {eqnarray}
\partial^{\lambda}F^{\mu\nu}+ \partial^{\nu}F^{\lambda\mu}+\partial^{\mu}F^{\nu\lambda}
                \hskip63mm&& \nonumber \\
             = \frac{1}{R} \left\{
                                -2 \left(\eta^{0\lambda}F^{\mu\nu} +
                                 \eta^{\mu 0}F^{\nu\lambda}+ \eta^{0\nu}F^{\lambda\mu}\right) \right.
                                 \hskip13mm&& \nonumber \\
                            \left.
                                 + x^{\alpha}\partial_{\alpha} \left( \eta^{0\lambda}F^{\mu\nu} +
                                 \eta^{\mu 0}F^{\nu\lambda}+ \eta^{0\nu}F^{\lambda\mu} \right) \right.
                                 \hskip8mm&& \nonumber \\
                            \left.
                                 + x^{\mu}\partial^{0}F^{\nu\lambda} + x^{\nu}\partial^{0}F^{\lambda\mu} +
                                 x^{\lambda}\partial^{0}F^{\mu\nu}
                            \right\}.
\end {eqnarray}
This equation represents the first group of Maxwell's equations, up to the first order,
in $R$-Minkowski spacetime.

Contracting two indices gives
\begin {eqnarray}
\partial_{\mu}F^{\mu\lambda}+ \partial_{\mu}F^{\lambda\mu}
             = \frac{1}{R} \left\{
                                -2 \left( F^{0\lambda}  + F^{\lambda 0} \right) \right.
                                \hskip47mm&& \nonumber \\
                                 \left. + x_{\mu}\left( \partial^{\mu}F^{0\lambda}+ \partial^{\mu}F^{\lambda 0} \right)
                                 + x_{\mu}\left( \partial^{0}F^{\mu\lambda}+ \partial^{0}F^{\lambda \mu} \right)
                            \right\}.
\end {eqnarray}
With an analogous procedure as in the context of special relativity \cite{BGM1} , let us define the four-current
in the following manner
\begin{equation}
\mu_{0}J^{\lambda} = - \partial_{\mu}F^{\lambda\mu}
                       + \frac{1}{R} \left(-2\xi F^{\lambda 0} + \sigma x_{\mu} \partial^{\mu}F^{\lambda 0}
                       + \omega x_{\mu} \partial^{0}F^{\lambda \mu}  \right),
\end{equation}
where the constants $\xi$, $\sigma$ and $\omega$ can take freely any value about $0$, $+1$ and $-1$.
Because of the antisymmetric feature of the tensor $F^{\mu\nu}$, relation (35) allows to write the
second group of Maxwell's equations as
\begin{equation}
 \partial_{\mu}F^{\mu\lambda}
             + \frac{1}{R} \left(2\xi F^{0 \lambda } - \sigma x_{\mu} \partial^{\mu}F^{0 \lambda}
                                  - \omega x_{\mu} \partial^{0}F^{\mu \lambda}  \right)
                                           = \mu_{0}J^{\lambda}.
\end{equation}
In order to determine the exact values of $\xi$, $\sigma$ and $\omega$, we will impose that the
well-known electric-magnetic duality be a symmetry of the above Maxwell's equations in the absence
of the sources. For this purpose, let us introduce the electric and the magnetic fields,
\begin{equation}
E^{i}=cF^{i0}, \hskip35mm B_{i}=\varepsilon_{ijk}F^{jk}/2,
\end{equation}
where $\varepsilon_{ijk}$ is the Levi-Civita
antisymmetric tensor $(\varepsilon_{123}=1)$. From (34) and (37), we can deduce that
\begin{align}
               \frac{1}{c} \overrightarrow{\nabla} \wedge \overrightarrow{E} + \partial_{0} \overrightarrow{B}
                + \frac{1}{R} \left[ 2\overrightarrow{B} - x^{0} \partial_{0} \overrightarrow{B} -
                         x^{\alpha}\partial_{\alpha} \overrightarrow{B} +
                         \frac{1}{c} \overrightarrow{r} \wedge \partial_{0} \overrightarrow{E}
                         \right] & = \overrightarrow{0} ,\\
               \overrightarrow{\nabla}\cdot\overrightarrow{B} +
                       \frac{1}{R} \overrightarrow{r}\cdot \partial_{0} \overrightarrow{B} & = 0 ,    \\
                \overrightarrow{\nabla}\cdot\overrightarrow{E} +
                       \frac{\omega}{R} \overrightarrow{r}\cdot \partial_{0} \overrightarrow{E} & =  c\mu_{0}{j^{0}} ,   \\
                \overrightarrow{\nabla} \wedge \overrightarrow{B} -\frac{1}{c} \partial_{0} \overrightarrow{E}
                 - \frac{1}{R} \left[ \frac{2\xi}{c}\overrightarrow{E} -
                                      \frac{\omega}{c} x^{0} \partial_{0} \overrightarrow{E}  \right. \hskip27mm&& \nonumber \\
                             \left. - \frac{\sigma}{c} x^{\alpha}\partial_{\alpha} \overrightarrow{E} -
                         \omega \overrightarrow{r} \wedge \partial_{0} \overrightarrow{B}
                         \right] & = \mu_{0}\overrightarrow{j}.
\end{align}
In the absence of the source, it is easy to show that the invariance of the above equations under the
electric-magnetic duality
\begin{equation}
\overrightarrow{E} \mapsto c\overrightarrow{B},   \ \ \ \ \ \ \ \ \ \ \ \
   \overrightarrow{B} \rightarrow - \frac{1}{c}\overrightarrow{E}
\end{equation}
requires to take $\xi=\sigma=\omega=+1$. With these values, (37) represents the second
group of Maxwell's equations in $R$-Minkowski spacetime, up to the first order.

Let us now move on to the Lorentz force. As in special relativity, the force is given by
$F^{\mu}=m\ddot{x}^{\mu}$ and then
\begin{equation}
\left[x^{\mu},F^{\nu} \right] = m \left[x^{\mu},\ddot{x}^{\nu}\right]
                              = m \frac{d}{d\tau}\left[x^{\mu},\dot{x}^{\nu} \right]
                                             - m \left[\dot{x}^{\mu},\dot{x}^{\nu} \right].
\end{equation}
Using (27) and (30), we obtain
\begin{equation}
\left[x^{\mu},F^{\nu} \right] = \frac{i\hbar q}{m} F^{\mu\nu} +
                                \frac{2i\hbar }{R} \left(\eta^{\mu\nu} \dot{x}^{0} + \eta^{\mu 0} \dot{x}^{\nu} \right).
\end{equation}
At zeroth order, we have
\begin{equation}
\frac{i\hbar q}{m} F^{\mu\nu} = \left[x^{\mu},F^{\nu} \right]_{(0)}
                              = \left[x^{\mu},\dot{x}^{\alpha}\right]_{(0)} \frac{\partial F^{\nu} }{\partial \dot{x}^{\alpha}}
                              = -\frac{i\hbar }{m} \frac{\partial F^{\nu} }{\partial \dot{x}_{\mu}},
\end{equation}
which gives
\begin{equation}
F^{\nu} = <q F^{\nu\mu} \dot{x}_{\mu} >+ G^{\nu}(x),
\end{equation}
where $G^{\nu}$ is an arbitrary function of $x$ and the symbol $<...>$ refers to symmetrization.
Thus, equation (45) indicates that the expression at the first order for $F^{\nu}$ will take the form
\begin{equation}
F^{\nu} =  <q F^{\nu\mu} \dot{x}_{\mu} > + G^{\nu}(x) + \frac{1}{R} H^{\nu}(x,\dot{x}),
\end{equation}
where $H^{\nu}$ is independent on $R$.
As $F^{\mu\nu}$ does not depend on $\dot{x}$, using this last expression of $F^{\nu}$ and taking into
account (27), we obtain
\begin{equation}
\left[x^{\mu},F^{\nu} \right] = \frac{i\hbar q}{m} F^{\mu\nu} +
                                \frac{i\hbar q}{mR} \left(2x^{0}F^{\nu\mu} + \eta^{\mu 0} x_{\lambda} F^{\nu\lambda}
                                                          + F^{\nu 0} x^{\mu} \right) +
                                \frac{1}{R} \left[x^{\mu},H^{\nu} \right].
\end{equation}
Identifying (45) and (49), we deduce
\begin{equation}
\left[x^{\mu},H^{\nu} \right] = 2i\hbar \left(\eta^{\mu\nu} \dot{x}^{0} + \eta^{\mu 0} \dot{x}^{\nu} \right) -
                               \frac{i\hbar q}{m} \left(2x^{0}F^{\nu\mu} + \eta^{\mu 0} x_{\lambda} F^{\nu\lambda}
                                                          + F^{\nu 0} x^{\mu} \right).
\end{equation}
In (49), the commutator $\left[x^{\mu},H^{\nu} \right]$ can be replaced by its value of zeroth order
since it is preceded by the factor $1/R$. At this order, we have
\begin{equation}
\left[ \dot{x}^{\mu},\left[x^{\nu},\dot{x}^{\lambda} \right]_{(0)} \right]_{(0)} = 0
\end{equation}
and therefore
\begin{equation}
\left[x^{\mu},H^{\nu} \right]_{(0)}
               = \left[x^{\mu},\dot{x}^{\lambda} \right]_{(0)}\frac{\partial H^{\nu}}{\partial \dot{x}^{\lambda}}
               = - \frac{i\hbar }{m} \frac{\partial H^{\nu}}{\partial \dot{x}_{\mu}}.
\end{equation}
With the use of (50), we find
\begin{equation}
\frac{\partial H^{\nu}}{\partial \dot{x}_{\mu}} = -2m \left(\eta^{\mu\nu} \dot{x}^{0} +
                \eta^{\mu 0} \dot{x}^{\nu} \right) + q \left(2x^{0}F^{\nu\mu} + \eta^{\mu 0} x_{\lambda} F^{\nu\lambda}
                                                          + F^{\nu 0} x^{\mu} \right),
\end{equation}
and we then deduce
\begin {eqnarray}
H^{\nu}(x, \dot{x})= - m \left(\dot{x}^{\nu} \dot{x}^{0}  + \dot{x}^{0} \dot{x}^{\nu}\right)  \hskip45mm&&  \nonumber \\
                   + < 2q x^{0}F^{\nu\mu} \dot{x}_{\mu}  +  q F^{\nu\lambda} x_{\lambda} \dot{x}_{0}
                   +  q F^{\nu 0} x^{\mu}\dot{x}_{\mu} >.
\end {eqnarray}
With this result, expression (48) turns out to be
\begin {eqnarray}
F^{\nu} = <q F^{\nu\mu} \dot{x}_{\mu} > + G^{\nu}(x)  +
        \frac{1}{R}  \left[ -m \left(\dot{x}^{\nu} \dot{x}^{0} + \dot{x}^{0} \dot{x}^{\nu}\right)  \right.
                            \hskip20mm&&  \nonumber \\
                            \left. +  < 2q x^{0}F^{\nu\mu} \dot{x}_{\mu} + q F^{\nu\lambda} x_{\lambda} \dot{x}_{0}
                            + q F^{\nu 0} x^{\mu}\dot{x}_{\mu} >
                    \right] ,
\end {eqnarray}
Contrary to the DSR case \cite{H1}, the Lorentz force in $R$-Minkowski spacetime contains
a corrective term proportional to $1/R$ at first order.

\section{The $R$-Lorentz symmetry}

Now, we will focus on some consequences of the above extension of Maxwell's equations on
the $R$-Lorentz symmetry.

First, we consider the $R$-Lorentz Lie algebra in the absence of the
electromagnetic field. In this case, we will note $J^{\mu\nu}$ the angular momentum components.
Multiplying at left equation (16) by $\left(1+x^{0}/R\right)$, and using successively (22) and (20),
we find at first order in $1/R$ the following expression for the momentum
\begin{equation}
p^{\mu} = m\dot{x}^{\mu} + \frac{m}{R} \left(x^{\mu}\dot{x}^{0} + 2 x^{0}\dot{x}^{\mu}\right).
\end{equation}
Using this expression in (5), the angular momentum becomes
\begin{equation}
J^{\mu\nu} = m\left( 1+ \frac{2x^{0}}{R}\right) \left(x^{\mu}\dot{x}^{\nu} - x^{\nu}\dot{x}^{\mu}) \right).
\end{equation}
The symmetrization operation allows us to write
\begin{eqnarray}
J^{\mu\nu} = \frac{m}{2} \left( x^{\mu}\dot{x}^{\nu} + \dot{x}^{\nu} x^{\mu}
                                - x^{\nu}\dot{x}^{\mu} - \dot{x}^{\mu} x^{\nu}\right)
            + \frac{m}{3R} \left(2x^{0}x^{\mu}\dot{x}^{\nu} + x^{0}\dot{x}^{\nu}x^{\mu}  \right.  \hskip8mm&&  \nonumber \\
                                 \left. + x^{\mu}\dot{x}^{\nu}x^{0} + 2\dot{x}^{\nu}x^{\mu}x^{0}
                                  - 2x^{0} x^{\nu}\dot{x}^{\mu} - x^{0} \dot{x}^{\mu}x^{\nu}
                                 - x^{\nu}\dot{x}^{\mu}x^{0} - 2\dot{x}^{\mu}x^{\nu}x^{0} \right).
\end {eqnarray}
From relation (27), it is easy to show that
$\left[x^{\mu},\dot{x}^{\nu} \right]= \left[x^{\nu},\dot{x}^{\mu} \right]$
and then to deduce that
\begin{equation}
x^{\mu}\dot{x}^{\nu} - x^{\nu}\dot{x}^{\mu} = \dot{x}^{\nu}x^{\mu} - \dot{x}^{\mu}x^{\nu}.
\end{equation}
It follows that
\begin{equation}
J^{\mu\nu} = m\left( x^{\mu}\dot{x}^{\nu} - x^{\nu}\dot{x}^{\mu}\right)
           + \frac{m}{R} \left[ x^{0}\left( x^{\mu}\dot{x}^{\nu} - x^{\nu}\dot{x}^{\mu}\right) +
             \left( x^{\mu}\dot{x}^{\nu} - x^{\nu}\dot{x}^{\mu}\right) x^{0} \right].
\end{equation}
Again, the use of (27) allows us to write
\begin{equation}
\dot{x}^{\mu} x^{0} = \frac{i\hbar}{m} \eta^{0\mu} -
                      \frac{i\hbar}{mR} \left(3x^{0}\eta^{0\mu} + x^{\mu} \right) + x^{0}\dot{x}^{\mu}
\end{equation}
with which the angular momentum takes finally the following simple form
\begin{equation}
J^{\mu\nu} = m\left(1+ \frac{2x^{0}}{R}  \right)\left( x^{\mu}\dot{x}^{\nu} - x^{\nu}\dot{x}^{\mu}\right)
           + \frac{i\hbar}{R} \left( x^{\mu} \eta^{0\nu} - x^{\nu} \eta^{0\mu}\right).
\end{equation}
Thus, with the help of (27) and (28), we can show after some calculations that
\begin{align}
\left[x^{\mu},J^{\nu\lambda} \right] & = i\hbar \left(\eta^{\mu\nu} x^{\lambda} - \eta^{\mu\lambda} x^{\nu} \right)
                                         + \frac{i\hbar}{R} \left( \eta^{0\lambda}x^{\mu}x^{\nu}
                                                               - \eta^{0\nu} x^{\mu}x^{\lambda}\right), \\
\left[\dot{x}^{\mu},J^{\nu\lambda} \right] & =  i\hbar \left(\eta^{\mu\nu} \dot{x}^{\lambda} - \eta^{\mu\lambda} \dot{x}^{\nu} \right)
                                                - \frac{i\hbar}{R} x^{\mu} \left(\eta^{0\nu} \dot{x}^{\lambda}
                                                - \eta^{0\lambda} \dot{x}^{\nu} \right) \nonumber \\
                                           &    \hskip4mm - \frac{i\hbar}{R} \left(\eta^{0\nu} x^{\lambda}
                                                - \eta^{0\lambda} x^{\nu} \right) \dot{x}^{\mu}
                                                - \frac{\hbar^{2}}{Rm} \left(\eta^{0\lambda}\eta^{\mu\nu}
                                                - \eta^{0\nu} \eta^{\mu\lambda} \right), \\
\left[J^{\mu\nu},J^{\lambda\sigma}\right] & =   i\hbar \left(\eta^{\nu\lambda}J^{\mu\sigma}  -  \eta^{\nu\sigma}J^{\mu\lambda}
                                                     - \eta^{\mu\lambda}J^{\nu\sigma} + \eta^{\mu\sigma}J^{\nu\lambda}\right).
\end{align}
We observe that the commutators (63) and (65) have the same forms as in (6) and (8) while (64) can
be derived from (7).

In the presence of the electromagnetic field, we will note $M^{\mu\nu}$ the contribution of the field to
the angular momentum and therefore we write
\begin{equation}
\textbf{J}^{\mu\nu} = J^{\mu\nu} + M^{\mu\nu}.
\end{equation}
Also, for the commutator $\left[\dot{x}^{\mu},\dot{x}^{\nu} \right]$, it is expression (30) which must be used
instead of (28). Then, the above $R$-Lorentz Lie algebra turns out to be
\begin{align}
[x^{\mu}, \textbf{J}^{\nu\lambda}] & = -i\hbar\left(\eta^{\mu\lambda}x^{\nu}-\eta^{\mu\nu}x^{\lambda}\right) +
                                      \frac{i\hbar}{R}x^{\mu}\left(\eta^{0\lambda}x^{\nu}-\eta^{0\nu}x^{\lambda}\right) +
                                      \left[x^{\mu},M^{\nu\lambda}\right] \\
\left[\dot{x}^{\mu},\textbf{J}^{\nu\lambda} \right] & =  i\hbar \left(\eta^{\mu\nu} \dot{x}^{\lambda} - \eta^{\mu\lambda} \dot{x}^{\nu} \right)
                                                - \frac{i\hbar}{R} x^{\mu} \left(\eta^{0\nu} \dot{x}^{\lambda}
                                                - \eta^{0\lambda} \dot{x}^{\nu} \right) \nonumber \\
                                           &    \hskip4mm - \frac{i\hbar}{R} \left(\eta^{0\nu} x^{\lambda}
                                                - \eta^{0\lambda} x^{\nu} \right) \dot{x}^{\mu}
                                                - \frac{\hbar^{2}}{Rm} \left(\eta^{0\lambda}\eta^{\mu\nu}
                                                - \eta^{0\nu} \eta^{\mu\lambda} \right) \nonumber \\
                                           &    \hskip4mm - \frac{i\hbar q}{m} \left(1+ \frac{2x^{0}}{R}\right)
                                                \left(x^{\nu}F^{\mu\lambda} - x^{\lambda}F^{\mu\nu}\right) +
                                                \left[\dot{x}^{\mu},M^{\nu\lambda}\right], \\
\left[\textbf{J}^{\mu\nu},\textbf{J}^{\lambda\sigma}\right] & =   i\hbar \left(\eta^{\nu\lambda}J^{\mu\sigma} -
                                                     \eta^{\nu\sigma}J^{\mu\lambda} - \eta^{\mu\lambda}J^{\nu\sigma}
                                                      + \eta^{\mu\sigma}J^{\nu\lambda}\right) \nonumber \\
                                          &    \hskip4mm + i\hbar q \left(1+ \frac{4x^{0}}{R}\right) \left(x^{\mu}x^{\sigma}F^{\nu\lambda}
                                                        - x^{\mu}x^{\lambda}F^{\nu\sigma} - x^{\nu}x^{\sigma}F^{\mu\lambda} \right. \nonumber \\
                                          &    \hskip4mm \left.  + x^{\nu}x^{\lambda}F^{\mu\sigma}\right) + \left[J^{\mu\nu}, M^{\lambda\sigma}\right]
                                                + \left[M^{\mu\nu}, J^{\lambda\sigma}\right] + \left[M^{\mu\nu}, M^{\lambda\sigma}\right].
\end{align}
In order to restore the $R$-algebra constituted by (6), (7) and (8) and represented
at first order in the absence of the electromagnetic field by (63), (64) and (65), we will impose:
\begin{align}
& [x^{\mu}, M^{\nu\lambda}] = 0,    \\
& [\dot{x}^{\mu}, M^{\nu\lambda}] -\frac{i\hbar q}{m}\left(1+\frac{2x^{0}}{R}\right) \left(x^{\nu}F^{\mu\lambda}-x^{\lambda}F^{\mu\nu}\right)  =  0,  \\
& i\hbar \left(-\eta^{\nu\lambda}M^{\mu\sigma} + \eta^{\nu\sigma}M^{\mu\lambda} +
                         \eta^{\mu\lambda}M^{\nu\sigma} - \eta^{\mu\sigma}M^{\nu\lambda}\right)      \nonumber \\
 &    \hskip2mm + i\hbar q \left(1+ \frac{4x^{0}}{R}\right) \left(x^{\mu}x^{\sigma}F^{\nu\lambda}
                                                        - x^{\mu}x^{\lambda}F^{\nu\sigma} - x^{\nu}x^{\sigma}F^{\mu\lambda}
                                            + x^{\nu}x^{\lambda}F^{\mu\sigma}\right) \hskip1mm      \nonumber \\
 &    \hskip34mm + \left[J^{\mu\nu}, M^{\lambda\sigma}\right] + \left[M^{\mu\nu}, J^{\lambda\sigma}\right] +
                                           \left[M^{\mu\nu}, M^{\lambda\sigma}\right]     = 0.
\end{align}
This last relation is obtained from (69) by substituting the components $J^{\mu\nu}$ by $\textbf{J}^{\mu\nu} - M^{\mu\nu}$
and requiring (8) to hold. Condition (70) implies that $M^{\mu\nu}$ does not depend on $\dot{x}$. It follows that the last
commutator in (72) takes a vanishing value. The two others can be determined by using expression (62) of $J^{\mu\nu}$
\begin{eqnarray}
\left[J^{\mu\nu}, M^{\lambda\sigma}\right] + \left[M^{\mu\nu}, J^{\lambda\sigma}\right] =
                                    m\left(1+ \frac{2x^{0}}{R}  \right)
                                    \left\{ x^{\mu}\left[\dot{x}^{\nu} , M^{\lambda\sigma}\right]
                                    \right. \hskip19mm&&  \nonumber \\
                                    \left. - x^{\nu}\left[\dot{x}^{\mu} , M^{\lambda\sigma}\right]
                                    - x^{\lambda}\left[\dot{x}^{\sigma} , M^{\mu\nu}\right]
                                    + x^{\sigma}\left[\dot{x}^{\lambda} , M^{\mu\nu}\right] \right\}.
\end{eqnarray}
Condition (71) determines the commutator $[\dot{x}^{\mu}, M^{\nu\lambda}]$ and allows therefore to write
\begin{eqnarray}
\left[J^{\mu\nu}, M^{\lambda\sigma}\right] + \left[M^{\mu\nu}, J^{\lambda\sigma}\right] =
                                    2i\hbar q \left(1+ \frac{4x^{0}}{R}  \right)
                                    \left[ x^{\mu}x^{\lambda}F^{\nu\sigma} \right. \hskip24mm&&  \nonumber \\
                                    \left. - x^{\mu}x^{\sigma}F^{\nu\lambda} - x^{\nu}x^{\lambda}F^{\mu\sigma}
                                    + x^{\nu}x^{\sigma}F^{\mu\lambda} \right].
\end{eqnarray}
Using this result in condition (72), we obtain
\begin{eqnarray}
 \eta^{\nu\lambda}M^{\mu\sigma} - \eta^{\nu\sigma}M^{\mu\lambda} -
                         \eta^{\mu\lambda}M^{\nu\sigma} + \eta^{\mu\sigma}M^{\nu\lambda}\hskip44mm &     \nonumber \\
                         =     q \left(1+ \frac{4x^{0}}{R}  \right)
                                    \left[ x^{\mu}x^{\lambda}F^{\nu\sigma}
                                     - x^{\mu}x^{\sigma}F^{\nu\lambda} - x^{\nu}x^{\lambda}F^{\mu\sigma}
                                    + x^{\nu}x^{\sigma}F^{\mu\lambda} \right].
\end{eqnarray}
Let us consider the spatial components by performing in the last relation the substitution
of the indices $(\mu, \nu, \lambda, \sigma)$ respectively by $(i,j,k,l)$.  Then, by contracting
the obtained equation with $\eta_{il}$, we get to
\begin{equation}
M^{jk}=q\left(1+\frac{4x^{0}}{R}\right)(x^{k}x_{l}F^{jl}-x_{l}x^{l}F^{jk}+x^{j}x_{l}F^{lk}),
\end{equation}
where we have used the fact that $M^{\mu\nu}$ is, from its definition (66), antisymmetric. Let us define
\begin{equation}
M_{i} \equiv \frac{1}{2} \varepsilon_{ijk} M^{jk} =
          \frac{1}{2}\varepsilon_{ijk} \ q\left(1+\frac{4x^{0}}{R}\right)(x^{k}x_{l}F^{jl}-x_{l}x^{l}F^{jk}+x^{j}x_{l}F^{lk}),
\end{equation}
$M_{i}$ is called the magnetic angular momentum \cite{BGM2}. From the second equation in (38), we have
\begin{equation}
F^{ij} = \varepsilon^{kij} B^{k}.
\end{equation}
The use of this relation in (77) leads to
\begin{equation}
M^{i}=q\left(1+\frac{4x^{0}}{R}\right)(x^{j}B^{j})x^{i},
\end{equation}
which can be written as
\begin{equation}
\overrightarrow{M}=q\left(1+\frac{4x^{0}}{R}\right)\left(\overrightarrow{r}\cdot\overrightarrow{B}\right)\overrightarrow{r}.
\end{equation}
Otherwise, as $M^{jk}$ does not depend on $\dot{x}$ and taking into account (27), we have
\begin{align}
[\dot{x}^{i}, M^{jk}] & = [\dot{x}^{i}, x^{\mu}] \frac{\partial M^{jk}}{\partial x^{\mu}} \nonumber \\
                      & = \frac{i\hbar}{m}\left(1-\frac{2x^{0}}{R}\right)\frac{\partial M^{jk}}{\partial x_{i}}
                          - \frac{i\hbar}{mR}\frac{\partial M^{jk}}{\partial x_{0}} x^{i}.
\end{align}
In the case where $\overrightarrow{B}$ does not depend on $x^{0}$, (76) indicates that the zeroth order
of $M^{jk}$ does not depend on $x^{0}$. Thus, taking into account condition (71), relation (81) allows
to write at first order
\begin{equation}
\frac{\partial M^{jk}}{\partial x_{i}}=q \left(1+\frac{4x^{0}}{R}\right)(x^{j}F^{ik}-x^{k}F^{ij}).
\end{equation}
By multiplying by $\varepsilon^{ljk} $ this last equation and using (77) and (78),
we obtain
\begin{equation}
\frac{\partial M^{l}}{\partial x_{i}}= q \left(1+\frac{4x^{0}}{R}\right)
                       \left(B^{l}x^{i} - \overrightarrow{r}\cdot \overrightarrow{B} \delta^{li}  \right).
\end{equation}
With the help of (79), this last relation gives
\begin{equation}
\frac{\partial B^{j}}{\partial x_{i}}x^{j}x_{l} = B^{i}x_{l} + B_{l}x^{i}.
\end{equation}
As in \cite{BGM2}, the solution of this equation is
\begin{equation}
B^{i} = \frac{\mu_{0} g}{4\pi}  \frac{x^{i}}{r^{3}},
\end{equation}
where $r=\left(x^{j}x^{j}\right)^{1/2}$ and $g$ is an integration constant. It is easy
to check that $\overrightarrow{\nabla}\cdot\overrightarrow{B}=\mu_{0}g\delta(\overrightarrow{r})$.
This means that $g$ can be identified to the magnetic charge and that equation (85) describes
the Dirac magnetic Monopole. Substituting this expression
of $B^{i}$ in (80), we obtain the magnetic angular momentum
\begin{equation}
\overrightarrow{M}=\frac{\mu_{0}qg}{4\pi}\left(1+\frac{4x^{0}}{R}\right) \frac{\overrightarrow{r}}{r}.
\end{equation}

We also observe that in a stationary case, the electric field keeps its usual form
without any corrective term. In fact, on the one hand, condition (71) allows to write
\begin{equation}
[\dot{x}^{0}, M^{ij}] = \frac{i\hbar q}{m}\left(1+\frac{2x^{0}}{R}\right) \left(x^{i}F^{0j}-x^{j}F^{0i}\right),
\end{equation}
and on the other, from (77) and (86), we obtain
\begin{equation}
M^{ij}=\varepsilon^{ijk} M^{k}=\frac{\mu_{0}qg}{4\pi}\varepsilon^{ijk} \left(1+\frac{4x^{0}}{R}\right) \frac{x^{k}}{r},
\end{equation}
with which we can deduce
\begin{equation}
[\dot{x}^{0}, M^{ij}] = [\dot{x}^{0}, x^{l}] \frac{\partial M^{ij}}{\partial x^{l}}=0.
\end{equation}
Relations (87) and (89) indicate that
\begin{equation}
x^{i}F^{0j}-x^{j}F^{0i} = 0,
\end{equation}
which leads to the following usual expression for the electric field :
\begin{equation}
E^{i} = Kq \frac{x^{i}}{r^{3}}.
\end{equation}
The last observation concerns the component $M^{0i}$. If we perform in (75) the substitution
of the indices $(\mu, \nu, \lambda, \sigma)$ respectively by $(0,j,i,j)$, we can show that $M^{0i}=0$.

\section{Conclusion}

Up to first order in the deformation, we have derived in $R$-Minkowski Spacetime the Maxwell Equations
by using a generalized version of Feynman's method. The electric-magnetic duality symmetry is imposed to
fix some arbitrary parameters appearing in the present approach. We have also established the expression
of the Lorentz force in this context. The $R$-Lorentz algebra, established in \cite{BF}, is restored in
the presence of the electromagnetic field by adding to the angular momentum  the electromagnetic field
contribution which generates the Dirac magnetic monopole. Contrary to the DSR case \cite{H1},
the angular magnetic momentum is affected by the parameter deformation. We would like to add that unlike
the earlier works in which the  form (85) of the Dirac magnetic monopole was obtained without
imposing the no dependence on the time despite its similarity with the electrostatic field,
in our present paper, this condition is necessary to reach expression (85).

\end{document}